\documentclass[aps,amsmath,amsfonts,twocolumn]{revtex4-2}%
\usepackage{mathtools}
\usepackage{verbatim}
\usepackage{xr-hyper}
\usepackage{hyperref}
\usepackage{cleveref}
\usepackage{epsfig,color,amsfonts,amsmath,amsthm,comment,natbib}
\usepackage{color}
\usepackage{graphicx}
\usepackage{braket}
\usepackage{amsthm}
\usepackage{amssymb}
\usepackage{amsmath}
\usepackage{ragged2e}
\usepackage{ulem}
\usepackage{tikz}
\usepackage[caption=false]{subfig}

\DeclareMathOperator{\Tr}{Tr}

\newcommand{\integers}{\mathbb{Z}}

\newcommand{\MW}{\color{blue}}
\newcommand{\old}{\color{black}}
\newcommand{\ee}{{\rm e}}
\newcommand{\ii}{{\rm i}}

\begin{document}

\title{Perfect State Transfer of Mixed States and Purification in Central Spin Systems}

\author{Matthew Wampler}
\author{Nigel R. Cooper}
\affiliation{T.C.M. Group, Cavendish Laboratory, University of Cambridge, J.J. Thomson Avenue, Cambridge CB3 0US, United Kingdom \looseness=-1}

\begin{abstract}
We show how two many-body, generally mixed, quantum states can be swapped via collective, all-to-all interactions. Specifically, we present an experimentally relevant implementation for quantum dots that enables coherent exchange of quantum information between different species of nuclear spins, effectively achieving a perfect swap gate for qudits formed from these different species. This process also serves as a tool for nuclear spin purification. The results are obtained by mapping the problem onto perfect state transfer on a 1D chain. In the partially polarized limit, we demonstrate that the states can be exchanged independently of the initial state. We also assess the robustness of the procedure to decoherence and errors.
\end{abstract}

\keywords{Generalized Dicke States}

\maketitle

\section{Introduction}
A major aim of 21st century physics has been the microscopic manipulation and control of quantum many-body systems for the purposes of engineering new quantum matter, applications in quantum information, and the investigation of non-equilibrium quantum physics more broadly. One form of dynamics that appears naturally in many such systems involves collective, all-to-all interactions.  Leveraging such interactions has become a powerful tool for control, used for efficient generation of many-body entanglement \cite{Molmer1999TrappedIons,Sorensen2000EntanglementGeneration,Hammerer2010AtomLightReview}, to achieve enhanced quantum sensing and metrology through spin-squeezing \cite{Leroux2010Squeezing,SchleierSmith2010Squeezing,Chen2011Squeezing}, investigate quantum phase transitions \cite{Hepp1973DickeTransition,Klinder2015OpenDicke,Muniz2020LMGTransition}, and enhance quantum memory as well as achieve efficient quantum state preparation \cite{Taylor2003CentralSpinMemory,Kalachev2006quantummemory,AsenjoGarcia2017QuantumMemory,Appel2025quantumregister,Yin2025GHZPrep}.  These benefits of global couplings are leveraged across a variety of physical platforms including cavity QED \cite{Ritsch2013cQED,Mivehvar2021cQED}, trapped ions \cite{FossFeig2025TrappedIonReview}, circuit QED \cite{Fink2009CirQED,Nissen2013CirQED,Armata2017CirQED}, and Central Spin systems \cite{Urbaszek2013CentralSpin,Taylor2003CentralSpin}.  

Abstractly, a general scenario which occurs in all such physical platforms is that there is some many-body system $A$ where the dynamics may be sufficiently well controlled (e.g. via driving and measurements) such that a desired behavior may be engineered.  This system, however, is coupled to an external system $B$ which is less well controlled; cross-talk between $A$ and $B$ may than act to ruin the desired dynamics in $A$.  Modeling the full system $A+B$ is complex and analysis is typically restricted to a weak coupling regime where system $B$ may be treated as a bath with its dynamics traced out.  Alternatively, one can consider scenarios where additional symmetries are present---such as permutation symmetry in all-to-all couplings between $A$ and $B$---which can allow for more tractable analysis \cite{Chase2008PermutationSymmetric}.  

In this work, we show how collective couplings between $A$ and $B$ can be engineered to swap the quantum states of the two subsystems, i.e. $\rho^A \otimes \rho^B \leftrightarrow \rho^B \otimes \rho^A$.  Furthermore, we will show that this strategy may be utilized to purify subsystem $B$ through its coupling to the controlled subsystem $A$.  
In particular, we will focus on a relevant implementation of this strategy in central spin systems.  

Central Spin systems, e.g. in quantum dots \cite{Urbaszek2013CentralSpin}, involve the collective coupling of a single electron spin to one or more species of nuclear spins.  Such systems have had a recent influx of interest due to rapidly advancing experimental control of the nuclear spins \cite{Urbaszek2013CentralSpin,daSilva2021strainfree,Jackson2022SpinPurification,Millington-Hotze2024polarizenuclearspins,Shofer2025tuningspininteraction}.  This newfound control may be leveraged to extend the coherence time of the central electron spin of the quantum dot as, when uncontrolled, the nuclear spins act as a main source of noise \cite{Bechtold2015spindecoherence,Malinowski2017spindecoherence}. Furthermore, new techniques have allowed for the storage and manipulation of quantum information in the nuclear spins themselves which have much longer coherence times than the central spin \cite{Taylor2003CentralSpinMemory,Zaporski2023preparesinglet,Appel2025quantumregister}.  Depending on the material system, there can be multiple, separately addressable nuclear spin species, thus offering the possibility for the electron spin to address multiple quantum registers.    

These strategies for using the nuclear spins as a many-body quantum register describe methods to transfer a quantum state from the electron spin to a single species of nuclear spin \cite{Song2005PSTCentralSpin,Appel2025quantumregister}.  In this work, we show that it is possible to leverage the collective all-to-all coupling between the different nuclear spin species---mediated by the central electron spin---to exchange quantum information between the nuclear spin species themselves.  

In order to do this, we find a mapping between the many-body problem of state exchange between nuclear spin species and perfect state transfer on a 1D chain.  Perfect state transfer is an operation which transports a state $|\psi\rangle$ from an initial site $0$ to a final site $N-1$ on a lattice of $N$ qubits, i.e. transfers the state $|\psi \rangle_0 |0\rangle^{\otimes N-1} \rightarrow |0\rangle^{\otimes N-1} |\psi \rangle_{N-1}$.  Such an operation---which is necessary to apply gates between distant qubits on quantum computing architectures with local interactions---typically requires applying a series of SWAP gates \cite{li2019tacklingqubitmappingproblem}.  An alternative was proposed in \cite{Bose2003PST}: engineer a Hamiltonian $H_{\rm PST}$ which acts on the entire system such that after a time $T$, 
\begin{gather}
\label{eq: perfect state transfer}
    \text{\underline{PST:} \;\;} |\langle 0|^{\otimes N-1} \langle \psi |_{N-1} \ee^{-\ii  H_{\rm PST} T}|\psi \rangle_0 |0\rangle^{\otimes N-1}|^2 = 1.
\end{gather}  
In theory, this could reduce errors as a single gate is achieving what would typically require an extensive number of swap gates (PST is also achievable in the case where the bulk of the qubit chain is left arbitrary \cite{Markiewicz2009PSTnomiddlecontrol}).  Furthermore, it has been shown that perfect state transfer may be used as a constructive tool to create more general quantum operations (see \cite{Kay2010PSTReview} for review).

Perfect state transfer has been discussed for mixed states as well as for open systems \cite{Wang2005transferwithmixed,Zhang2012PTSymTransfer,Gavreev2023SuppressDecoherencePST,Markiewicz2010NoisyPST,Hu2009PSTDecoherence,Ren2019PSTDecoherence,zhou2006decoherenceproblemquantumstate}.  Such dynamics is a special case of quantum walks in the presence of decoherence \cite{KENDON2007DecoherenceQuantumWalks}.  We show that, in the limit that two nuclear spin species $A$ and $B$ are partially polarized and have similar total spin, it is possible to achieve exchange of the quantum states independently of the initial---in general, mixed---states $\rho^A$ and $\rho^B$.  We additionally discuss the robustness of our procedure to decoherence.    

The rest of the paper is structured as follows.  In Section \ref{Sec: System Hamiltonian and Symmetries} we introduce the system Hamiltonian, symmetries of the model, as well as provide a simple example of purification in our set-up to illustrate the protocol.  Section \ref{Sec: PST Review} provides a brief review of the required machinery to achieve PST, albeit applied to a problem more relevant to central spin systems: magnetization inversion.  Section \ref{Sec: Mixed State Transfer} introduces the procedure for mixed state transfer between subsystems $A$ and $B$.  In Section \ref{Sec: Decoherence} we discuss robustness to decoherence and errors.  Final conclusions and a discussion are then provided in Section \ref{Sec: Conclusion}.

\section{System Hamiltonian and Symmetries}
\label{Sec: System Hamiltonian and Symmetries}
\begin{figure*}
    \centering
    \includegraphics[width=0.99\linewidth]{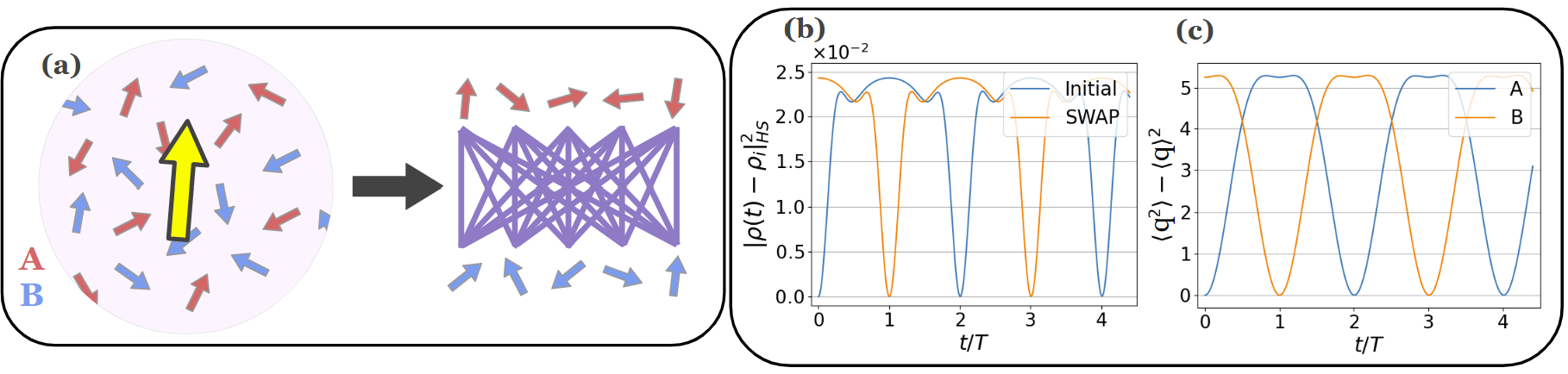}
    \caption{(a) Central Spin System: both red and blue nuclear spins are collectively coupled to the (yellow) electron spin.  When the electron spin is fixed, this leads to an effective all-to-all coupling between the two species of nuclear spins.  (b) Mixed State SWAP: an initial state $\rho_\text{initial}  \leftrightarrow \rho_\text{SWAP}$ is swapped under the system dynamics.  Subsystem $A$ is initialized in a pure state while $B$ is highly mixed in $J^B$ and $M^B$ (specific states and system parameters given in Sec. \ref{Sec: System Hamiltonian and Symmetries}).  The vanishing of the (weighted) Hilbert-Schmidt norm of the difference between the state of the whole system $\rho (t)$ and $\rho_i$ (corresponding to either the initial or swapped configuration) shows that the state of subsystems $A$ and $B$ are exchanged at the period $T$ found analytically in Sec. \ref{Sec: Purification Through Swapping}.  (c) The state of subsystem $B$ is purified at odd multiples of $T$ as the states are in the swapped configuration.  Here, $\langle q\rangle$ corresponds to the expectation value of the state of the qudit on subsystem $A$ (blue curve) or subsystem $B$ (orange curve).  Similarly, $\langle q^2\rangle$ is the expectation of the square of the qudit value.}
    \label{fig: central spin model and purity pumping}
\end{figure*}

We consider a quantum dot, central spin model \cite{Urbaszek2013CentralSpin}.  Here, a single central electron spin interacts with several species of surrounding nuclear spins, Fig \ref{fig: central spin model and purity pumping}(a).  In the dense limit, the individual nuclear spins cannot be distinguished by the electron spin and therefore any Hamiltonian will be symmetric under permutations of the nuclear spins.  It is therefore convenient to work with collective (total) spin component operators for each nuclear species ${\cal M}$---$J^{\cal M}_x,J^{\cal M}_y,J^{\cal M}_z$ where $J^{\cal M}_x=\frac{1}{2}\sum_{i\in {\cal M}} X_i$ for Pauli $X_i$ on nuclear spin $i$ in nuclear spin species $\cal M$ (units $\hbar=1$) and similarly for $y$,$z$---as they are generators of all permutation symmetric coherent evolution within species $\cal M$.  Furthermore, total spin is conserved under each of the total spin component operators, i.e. $\left[(\vec{J}^{\cal M})^2,J^{\cal M}_\alpha \right]=0$ for $\alpha \in \{x,y,z\}$ and $(\vec{J}^{\cal M})^2=\left(\sum_{i \in {\cal M}} \vec{J}_i\right)^2$ where $\vec{J}_i$ is the spin operator for the $i$th nuclear spin of species ${\cal M}$.  We will make frequent use of the basis diagonal in $(\vec{J}^{\cal M})^2$ and $J^{\cal M}_z$ 

\begin{gather}
    (\vec{J}^{\cal M})^2 |J,M,\xi \rangle_{\cal M} = J (J+1) |J,M,\xi \rangle_{\cal M}, \\
    J^{\cal M}_z |J,M,\xi \rangle_{\cal M} = M |J,M,\xi \rangle_{\cal M}.
\end{gather}
In the above, $\xi \in \{1,2,\ldots,d_{J}\}$ enumerates states degenerate in $(\vec{J}^{\cal M})^2$ and $J^{\cal M}_z$ with multiplicity $d_{J}$.  For brevity, we will also sometimes neglect writing $J$, $\xi$, i.e. $|J,M,\xi \rangle_{\cal M} \equiv |M\rangle_{\cal M}$, when they are conserved under the dynamics; this is the case for collective Hamiltonian evolution (constructed out of $J^{\cal M}_x,J^{\cal M}_y,J^{\cal M}_z$ terms).  It will also sometimes be necessary to differentiate between the quantum numbers $J,M,\xi$ of several different species, in which case we will write $J^{\cal M},M^{\cal M},\xi^{\cal M}$.

We may also interpret the magnetization of each nuclear spin species as a qudit, i.e. (writing in a more familiar way) we have 
\begin{gather}
    |M=-J+q \rangle_{\cal M} \equiv |q\rangle_{J^{\cal M}}
    \label{eq: qudit form}
\end{gather}
where $q \in \{0,1,\ldots,d-1 \}$ with $d=2J+1$.  Operations between qudits are then engineered via collective Hamiltonians which preserve $J^{\cal M},\xi^{\cal M} \; \forall \; {\cal M}$.  For convenience, we also define $\rho^{\cal M} (J^{\cal M}) = \sum_{q,q'} \rho^{\cal M}_{qq'} |q\rangle_{J^{\cal M}} \langle q'|_{J^{\cal M}}$ to be a generic density matrix for the state of the ${\cal M}$ qudit with total spin $J^{\cal M}$. 

In this work, we show that for a broad class of initial states $\rho^A(J^A) \otimes \rho^B(J^B)$ for two species $A,B$ it is possible to engineer a collective Hamiltonian $\cal H$ such that, at a specific time $T$, the states of the systems will have swapped.  Namely, we find
\begin{gather}
\label{eq: rho swap}
    \ee^{-\ii  {\cal H}T} \left[\rho^A(J^A) \otimes \rho^B(J^B) \right]\ee^{\ii {\cal H}T} = \rho^B(J^A) \otimes \rho^A(J^B).
\end{gather}
This can be true even when $\rho^A$ and $\rho^B$ have differing purity.  In this way, the swap may be viewed as a method to purify one species at the cost of depurifying the other.  Furthermore, swapping $A \leftrightarrow B$ continues to hold if the initial states are mixtures of $J^A$ or $J^B$, i.e. $\rho^{\cal M} \rightarrow \sum_{J^{\cal M}} \rho^{\cal M}(J^{\cal M})$, or if there is entanglement between the two species. 

The Hamiltonian we use to achieve this swap between two species of nuclear spins is given by
\begin{gather}
\label{eq: System Hamiltonian}
    {\cal H} = {\cal H}_\text{int} + {\cal H}_\text{tune}
\end{gather}
with
\begin{gather}
    \label{eq: Hint}
    {\cal H}_\text{int} = \gamma_\text{int} \left[J^A_+ J^B_- + J^A_- J^B_+\right], \\
    {\cal H}_\text{tune} = \left\{\gamma_M (J_Z^A + J_Z^B) + \gamma_J (\vec{J}^B)^{2}   \right\}\left[J^A_+ J^B_- + J^A_- J^B_+ \right] \label{eq: Htune}
\end{gather}
where $J^{\cal M}_\pm = J^{\cal M}_x \mp \ii J^{\cal M}_y$ are the spin raising and lowering operators, and $\gamma_\text{int},\gamma_M,\gamma_J$ are experimental parameters for the strengths of the Hamiltonian terms which will be tuned to achieve the swap.  For quantum dots, the terms in the Hamiltonian \eqref{eq: System Hamiltonian} can emerge, for example, as effective interactions through the hyperfine coupling between the central spin and two nuclear spin species in the limit of a large applied magnetic field \cite{Coish2008dotHamiltonian,Cywinski2009dotHamiltonian}.
This is effectively a perturbative expansion in the ratio of the precession frequency of the electron spin in the field of the nuclei to its precession frequency in the Zeeman field.  Such perturbative terms have been used in other contexts for control of the nuclear spin state, for example in \cite{Zaporski2023preparesinglet} where a $\pi$-gate for evolution with \eqref{eq: Hint} was proposed as part of a protocol for preparation of a many-body singlet state between two species of nuclear spins.    

In particular, we show that Eq. \eqref{eq: rho swap} may be achieved so long as two conditions are satisfied.  First, $A$ and $B$ must be partially polarized (without loss of generality, in the $-\hat{Z}$ direction).  In terms of the qudit basis, this corresponds to the initial state coefficients $\rho^A_{aa'} \rho^B_{bb'}$ only being non-zero if $\frac{a+b}{J^A+J^B},\frac{a'+b'}{J^A+J^B}\ll1$.  Another way to view this restriction is that it requires us to truncate the qudits such that $d\ll2J^{\cal M}+1$.  Second, we require $J^A \approx J^B$ in the sense that $J^B=J^A+\Delta$ where $|\frac{\Delta}{J^A}|\ll1$ (but where the initial state has definite total spin $J^A$ in the A subspace, swapping is still approximately achieved for mixed $J^A$ so long as the partially polarized condition is maintained).  If these two constraints are satisfied, then the state in subsystems $A$ and $B$ will be swapped for arbitrary mixtures of initial $J^B,M^A,M^B$.        

As an example to illustrate this effect, we choose as an initial (and swapped) state
\begin{equation}
\begin{aligned}
    \label{eq: initial rho}
    \rho_\text{initial} &= \frac{1}{N_{\uparrow,\text{max}} (2 \Delta_\text{max}+1)}|1 \rangle_{J^A} \langle 1 |_{J^A} \\
    &\otimes \sum_{J^B=J^A-\Delta_\text{max}}^{J^A + \Delta_\text{max}} \sum_{b=0}^{N_{\uparrow,\text{max}}-1} |b \rangle_{J^B} \langle b|_{J^B},
\end{aligned}
\end{equation}
\begin{equation}
\begin{aligned}
    \label{eq: swap of initial rho}
    \rho_\text{SWAP} &= \frac{1}{N_{\uparrow,\text{max}} (2 \Delta_\text{max}+1)} \sum_{a=0}^{N_{\uparrow,\text{max}}-1} |a \rangle_{J^A} \langle a |_{J^A} \\
    &\otimes \sum_{J^B=J^A-\Delta_\text{max}}^{J^A+\Delta_\text{max}} |1 \rangle_{J^B} \langle 1|_{J^B},
\end{aligned}
\end{equation}
with $J^A=50$, $\Delta_\text{max}=4$, and $N_{\uparrow,\text{max}}=8$.  Subsystem $A$ is initially in a pure state while $B$ is a mixture of different magnetizations and total spin.  Note, in this case, the conditions for swapping are satisfied as for each state in the mixture \eqref{eq: initial rho} we have both that $\frac{a+b}{J^A+J^B} \leq \frac{N_{\uparrow,\text{max}}}{2J^A-\Delta_\text{max}} = \frac{1}{12}$ and $|\frac{\Delta}{J^A}|\leq |\frac{\Delta_\text{max}}{J^A}| = \frac{2}{25}$.  The system parameters and time $T$ which achieve swapping are functions of $J^A$ and provided in Sec. \ref{Sec: Mixed State Transfer}.  In Fig. \ref{fig: central spin model and purity pumping}, it is shown that the square of the (weighted) Hilbert-Schmidt norm $|\rho(t)-\rho_i|_\text{HS}^2$ (where $\rho_i$ corresponds to either $\rho_\text{initial}$ or $\rho_\text{SWAP}$, see Appendix \ref{Appendix: HS norm} for discussion of the choice of norm) of the difference between the evolution of the density matrix of the full system $\rho(t)=\ee^{-\ii  {\cal H}t} \rho \ee^{\ii {\cal H}t}$ and the initial state vanishes at even multiples of $T$ whereas for the swapped configuration the norm vanishes at odd multiples of $T$.  This verifies that subsystems $A$ and $B$ swap at period $T$.  As a consequence, Fig. \ref{fig: central spin model and purity pumping}c shows that the qudit state of subsystem $B$ is purified at odd multiples of $T$.  Note, it is only the qudit state---and not any of the other quantum numbers---which evolve under the Hamiltonian.  For example, while under the procedure the qudit state of $B$ evolves from a mixture to a definite state, other quantum numbers (such as $J^B$) remain fixed at their initial mixture.   

\section{Review: Perfect State Transfer and Magnetization Inversion}
\label{Sec: PST Review}
Instead of focusing directly on the problem of PST, we will instead consider an identical problem more relevant to the all-to-all central spin systems in which we are interested in this work.  Namely, consider the problem of magnetization inversion (MI) of a pure state $|J,M,\xi \rangle \rightarrow |J,-M,\xi \rangle$ (we will only be working with a single nuclear spin species in this section, so drop the ${\cal M}$ label for brevity).  We are interested in all-to-all ($J$ and $\xi$ conserving) Hamiltonians $H_\text{MI}$  which achieve magnetization inversion at a finite time $T$, i.e. writing in the same form as \eqref{eq: perfect state transfer} we want $H_\text{MI}$ such that
\begin{gather}
\label{eq: Magenetization inversion}
    \text{\underline{MI:} \;\;} |\langle -M| \ee^{-\ii  H_\text{MI} T}|M\rangle|^2 = 1.
\end{gather}
For simplicity, we will focus on the case that $H_\text{MI}$ is of the form

\begin{equation}
\label{eq: H for pure magnetization inversion}
\begin{gathered}
    H_\text{MI} = \sum_{M=-J}^J V_M |M \rangle \langle M| \\
    + \sum_{M=-J}^{J-1} C_M (|M\rangle \langle M+1| + |M+1 \rangle \langle M|)
\end{gathered}
\end{equation}
with free parameters $C_M$, $V_M$.

The necessary and sufficient conditions for $H_\text{MI}$ to achieve MI at finite $T$ are given by (for proof, see \cite{Christandl2005PST,Shi2005PST,Kay2010PSTReview} and discussion below)
\\ \\
\underline{Conditions for MI:}
\begin{enumerate}
\label{list: Conditions for MI}
    \item Mirror symmetry $P$: $\left[H_\text{MI},P \right]=0$.
    \item Odd Commensurate Eigenvalues: \\$\lambda_{a+1}-\lambda_a = (2 m_a+1)\frac{\pi}{T}$.
\end{enumerate}
where the mirror symmetry operator $P=\prod_i X_i$ reflects the magnetization $P|M\rangle = |-M \rangle$, $\lambda_a$ is any eigenvalue in the spectrum of $H_\text{MI} |\lambda_a\rangle = \lambda_a |\lambda_a\rangle$ ordered such that $\lambda_a <\lambda_{a+1}$, $m_a \in \integers$, and $T$ is any real constant independent of $a$.  

We show in Appendix \ref{Appendix: Necessary and Sufficient Conditions} that the time it takes to achieve MI via evolving the system  under $H_\text{MI}$ is given precisely by $T$ from the odd commensurability condition.  The minimal time $T_\text{min}$ which satisfies the odd commensurate eigenvalue condition we will refer to as the swapping period.  In order to find the swapping period, note that taking any $T$ which is a solution to the odd commensurability condition and dividing by the greatest common divisor of the odd integers $(2 m_a + 1)$ will also be a valid $T$, i.e. $T_\text{min} = \frac{T}{\text{gcd}_a(2m_a+1 )}$.  Henceforth, we reserve the letter $T$ to refer to the swapping period, $T_\text{min} \rightarrow T$.     

The problem of MI \eqref{eq: Magenetization inversion} for a Hamiltonian of the form \eqref{eq: H for pure magnetization inversion} is isomorphic to the problem of PST on an inhomogeneous XXZ spin chain; discussion and proof of the necessary and sufficient conditions for PST in this case may be found in \cite{Christandl2005PST,Shi2005PST,Kay2010PSTReview}.  Proof of the conditions also therefore applies to the constraints on $H_\text{MI}$ for MI given above, however we review proof of these conditions in the context of MI in Appendix \ref{Appendix: Necessary and Sufficient Conditions} for completeness.  Intuitively, commensurability of the eigenvalues guarantees that the evolution is periodic.  Odd commensurability combined with mirror symmetry then guarantees that the system will evolve to the mirror reflected state after time $T$.  For discussion of conditions for PST with Hamiltonians of a more general form than \eqref{eq: H for pure magnetization inversion}, see \cite{Kay2010PSTReview} and references therein.    

The first condition for MI gives explicit constraints on the free parameters, namely for the Hamiltonian \eqref{eq: H for pure magnetization inversion} to be mirror symmetric we must have $C_M = C_{-M-1}$ and $V_M = V_{-M}$.  The second MI condition, however, is a condition on the eigenvalues.  Fulfilling this constraint is then an inverse eigenvalue problem, i.e. one must start with eigenvalues which fulfill the commensurate condition and from them construct the corresponding values of $C_M$, $V_M$ which yield the given spectrum.  Inverse eigenvalue problems are well studied and polynomial time algorithms have been found for tridiagonal matrices of the form \eqref{eq: H for pure magnetization inversion} as well as a variety of other relevant geometries \cite{Gladwell2005InverseProblems}.  

In addition to finding Hamiltonians for MI through numerical solution of the inverse eigenvalue problem, it is helpful to gain intuition by directly constructing Hamiltonians using physically relevant operators already known to satisfy the requirements for MI.  Note that each total spin component operator has eigenvalues given by $\{-J,-J + 1,\ldots, J \}$ and therefore satisfies the odd commensurate eigenvalue condition.  The operator $J_x$ fulfills the mirror symmetry condition as well.  In terms of the form \eqref{eq: H for pure magnetization inversion} we have
\begin{gather}
\label{eq: Jx coefficients for HMI}
    H_\text{MI}=\gamma_\text{MI}J_x: \;C_M = \gamma_\text{MI}\sqrt{(J-M)(J+M+1)}, \; V_M=0.
\end{gather}

To find the swapping period in the case of $H_\text{MI}=\gamma_\text{MI}J_x$, we have that $\lambda_{a+1}-\lambda_a = \gamma_\text{MI}$ and therefore
\begin{gather}
    T = \frac{\pi}{\gamma_\text{MI}}.
\end{gather}
We note that the independence of $T$ with respect to $J$, $\xi$, and initial $M$ also allows (trivially) for MI of mixed states.  Namely, a generic mixture of $|J,M,\xi \rangle$ states
\begin{gather}
    \rho_\text{MI}(J,M,\xi) = \sum_{J,M,\xi} \rho_{J,M,\xi} |J,M,\xi \rangle \langle J,M,\xi |
\end{gather}
will exhibit MI at period $T=\frac{\pi}{\gamma_\text{MI}}$,
\begin{gather}
    \ee^{-\ii  J_x \pi} \rho_\text{MI}(J,M,\xi) \ee^{\ii J_x \pi} = \rho_\text{MI}(J,-M,\xi),
\end{gather}
as each state in the mixture individually exhibits MI at that period.

In this special case, the result for MI could have been found very simply by noting that $\ee^{-\ii  J_x \pi} = \prod_a \ee^{-\ii  X_a \frac{\pi}{2}} = (-i)^N P$.  However, the framework reviewed in this section allows for more general $H_\text{MI}$ to be found.  Additionally, it provides a starting point for achieving operations beyond MI, such as mixed state transfer which we discuss now.   

\section{Mixed State Transfer}
\label{Sec: Mixed State Transfer}
In order to see how mixed state transfer---Equation \eqref{eq: rho swap}---is connected to perfect state transfer, it is helpful to work in the doubled Hilbert space.  Namely, we vectorize such that $\rho_{AB} \equiv \sum_{iji'j'}\rho^A_{i,j} \rho^B_{i',j'} |i\rangle\langle j| \otimes |i'\rangle\langle j'| \rightarrow \sum_{iji'j'}\rho^A_{i,j} \rho^B_{i',j'} |i\rangle | j\rangle |i'\rangle | j'\rangle \equiv |\rho_{AB} \rangle$.  Equation \eqref{eq: rho swap} then becomes

\begin{gather}
\label{eq: mixed PST}
    \ee^{-\ii  {\cal H}_\text{eff}T}|\rho_{AB} \rangle = |\rho_{BA}\rangle
\end{gather}
where ${\cal H}_\text{eff} = \frac{\ii}{T}\log \left[\ee^{-\ii  {\cal H}T}\otimes \ee^{-\ii  {\cal H}T}\right]$.

Achieving Eqn. \eqref{eq: mixed PST} is now manifestly a perfect state transfer problem.  Namely, we want to find a finite $T$ such that $\ee^{-\ii  {\cal H}_\text{eff}T} = P$ where $P$ is now a reflection of $A \leftrightarrow B$, i.e. $P |\rho_{AB} \rangle = |\rho_{BA} \rangle$.    

We will find that, in the case of our central spin Hamiltonian \eqref{eq: System Hamiltonian}, the conditions for mixed state transfer are identical to those for MI except with the new symmetry $P$.  In order to see this, we will need to specialize \eqref{eq: mixed PST} to central spin systems by utilizing the permutation symmetry of the nuclear spins.

\subsection{Permutation Symmetric Mixed States}
\label{Sec: Permutation Symmetric Mixed States}
The permutation symmetry of all operations acting on each nuclear spin species implies that the dynamics may be split into a number of symmetry sectors that grows exponentially with system size; it is helpful to work in a basis of mixed states which respect these symmetry sectors \cite{Chase2008PermutationSymmetric}.  This basis is closed under permutation symmetric Lindbladian evolution (and more general open dynamics).  Furthermore, the size of the basis within each symmetry sector scales only polynomially in system size as opposed to the exponential scaling of a general density matrix.  This, along with the algebraic structure of operations acting on this space, makes such systems much more amenable to both numerical and analytical investigation.      

For each nuclear spin species, it is convenient to construct this basis from the simultaneous eigenstates of $(\vec{J}^{\cal M})^2$,$J^{\cal M}_z$.  A general such density matrix $\rho^{\cal M}$ is given by (dropping $\cal M$ labels for brevity)

\begin{gather}
    \rho = \sum_{J} \sum_{M,M'} \rho_{J,M,M'} \left(\frac{1}{d_{J}}\sum_{\xi} \alpha_{\xi}|J,M,\xi \rangle \langle J,M',\xi|\right)\\
    \coloneqq \sum_{J} \sum_{M,M'} \rho_{J,M,M'} \overline{|J,M \rangle \langle J,M'|}.
    \label{eq: permutation symmetric density matrix}
\end{gather}
where the overbar is used to denote the sum over $\xi$ in the last line, $d_{J}$ is the multiplicity for states with total spin $J$ and any magnetization $M$, and $\alpha_{\xi}$ are constants fixed by initial conditions of the dynamics.  Any permutation symmetric open dynamics preserves $\xi$; this is the reason the initial mixture of $\xi$---set by $\alpha_{\xi}$---is fixed throughout the dynamics (thereby fixing an exponential number of degrees of freedom). The only relevant degrees of freedom for the dynamics are therefore $J$,$M$, and $M'$; each of these scale with system size, which implies it takes $O(N^3)$ coefficients $\rho_{J,M,M'}$ to characterize the state \eqref{eq: permutation symmetric density matrix} (as compared with the exponential number of parameters required for the full density matrix).    

Any (Hamiltonian or open) evolution generated by $J_x$, $J_y$, and $J_z$ will furthermore preserve $J$ leaving only evolution in the $M$,$M'$ sector.  We note, on the other hand, that permutation symmetric open dynamics which cannot be generated by total spin component operators (for example, symmetric single spin decay $\dot{\rho} = \sum_{i} \sigma_i^- \rho \sigma_i^+ - \frac{1}{2} \left\{\sigma_i^+ \sigma_i^-,\rho \right\}$) \textit{does} incoherently evolve $J$.  The general form of the evolution of \eqref{eq: permutation symmetric density matrix} under permutation symmetric open dynamics is given in \cite{Chase2008PermutationSymmetric}, which will be relevant when we consider decoherence in Section \ref{Sec: Decoherence}.

\subsection{Density Matrix Swapping}
\subsubsection{Purity Preserving Transfer when $\gamma_M=\gamma_J=0$}
\label{Sec: Swap with just Hint}
As a first step, we will consider the case of unitary evolution under ${\cal H}_\text{int}$ alone (i.e. $\gamma_M = \gamma_J=0$) and assume that both $\rho^A$ and $\rho^B$ start in a state with definite initial magnetization and total spin, i.e. $\rho^A_{\text{initial}} = \overline{|J^A,M^A\rangle \langle J^A, M^A|}$ and $\rho^B_{\text{initial}} = \overline{|J^B,M^B\rangle \langle J^B,M^B|}$.  We note that the collective dynamics \eqref{eq: System Hamiltonian} preserves $J^A$ and $J^B$, so we suppress writing them throughout this section.  We, furthermore, note that the system Hamiltonian  preserves total magnetization $M_\text{total} = M^A + M^B$.  Since the system Hamiltonian leaves $\xi^{\cal M}$ invariant, each state in the mixture $\frac{1}{d_{J^{\cal M}}}\sum_{\xi^{\cal M}} |J^{\cal M},M^{\cal M},\xi^{\cal M} \rangle \langle J^{\cal M},M'^{\cal M},\xi^{\cal M}|$ transforms in the same way under ${\cal H}$ and---to show perfect state transfer---it is thus sufficient to show PST for any individual pure state $|J^A,M^A,\xi^A\rangle |J^B,M^B,\xi^B\rangle \rightarrow |J^A,M^B,\xi^A\rangle |J^B,M^A,\xi^B\rangle$.  We here solve the now \textit{pure} perfect state transfer problem; in Appendix \ref{Appendix: mixed computational details} we solve the full mixed problem in the language of Eq. \eqref{eq: mixed PST} to show how it simplifies in this case to pure state transfer.  

We define a basis of pure states (with fixed $J^{\cal M},\xi^{\cal M},\xi^{\cal M'}$) in the $M_\text{total}$ symmetry sector
\begin{gather}
\label{eq: pure state transfer basis}
    |x\rangle = |M^A =-J^A+x\rangle |M^B=-J^B+N_\uparrow - x\rangle \\
    = |x \rangle_{J^A} |N_\uparrow-x \rangle_{J^B} \label{eq: qudit pure state transfer basis}
\end{gather}
where $N_\uparrow = J^A+J^B + M_\text{total}$, we have parameterized our states in terms of $x$ with $x \in \{0,1,\ldots,N_\uparrow \}$, and where we have written $|x \rangle$ both in terms of the magnetizations \eqref{eq: pure state transfer basis} and in terms of the corresponding qudit states \eqref{eq: qudit pure state transfer basis} for clarity.  In this basis, perfect state transfer is given by the transformation $x \leftrightarrow N_\uparrow-x$.  

The Hamiltonian \eqref{eq: System Hamiltonian} in the basis \eqref{eq: pure state transfer basis} (with $\gamma_M=\gamma_J=0$) is given by
\begin{gather}
\label{eq: H in pure basis}
    {\cal H} = \gamma_\text{int}\sum_x C_x \left(|x\rangle \langle x+1| + |x+1 \rangle \langle x| \right),\\
    C_x = \sqrt{(x+1)(2J^A-x)(N_\uparrow-x)(2J^B-N_\uparrow+x+1)}.
    \label{eq: Cx from H in pure basis}
\end{gather}
Note that this Hamiltonian is of the form \eqref{eq: H for pure magnetization inversion}.  In a similar fashion to the MI case, for $\cal H$ to achieve pure state transfer we must then have that $\left[{\cal H},P \right]=0$ and $\cal H$ has odd commensurate eigenvalues.

Here, the Hamiltonian is only symmetric if $J^A=J^B$ and the eigenvalues are not, in general, commensurate.  However, taking the partially polarized ($\frac{N_\uparrow}{J^A+J^B}\ll1$) and $J^A$ close to $J^B$ (i.e. $\frac{J^B-J^A}{J^A} \equiv \frac{\Delta}{J^A}\ll1)$ limits, $C_x$ becomes
\begin{equation}
\begin{gathered}
\label{eq: Cx in limits no tune}
    C_x = J \left(2 + \frac{\Delta}{J} - \frac{N_\uparrow-1}{2J} \right)\sqrt{(x+1)(N_\uparrow-x)}  \\
    + \text{ higher order terms}.
\end{gathered}
\end{equation}
where from now on we set $J^A\equiv J$ for brevity.

Note, with $C_x$ in the form \eqref{eq: Cx in limits no tune}, the Hamiltonian \eqref{eq: H in pure basis} satisfies the symmetry constraint for PST.  We now must check the odd commensurate eigenvalue constraint.  Since $J \left(2 + \frac{\Delta}{J} - \frac{N_\uparrow-1}{2J} \right)$ is independent of $x$, it may be pulled outside the sum over $x$ in \eqref{eq: H in pure basis} and acts simply to rescale $\gamma_\text{int}$.  Then \eqref{eq: H in pure basis} becomes exactly the same form as \eqref{eq: Jx coefficients for HMI} and therefore has spectrum $2J \left(2 + \frac{\Delta}{J} - \frac{N_\uparrow-1}{2J} \right) \left\{-\frac{N_\uparrow}{2},-\frac{N_\uparrow}{2}+1,\ldots,\frac{N_\uparrow}{2} \right\}$.  Hence, we have that the eigenvalues are odd commensurate and perfect state transfer will occur with period 
\begin{gather}
\label{eq: purity preserving period}
    T=\frac{\pi }{2 \gamma_\text{int} J} \frac{1}{\left(2 + \frac{\Delta}{J} - \frac{N_\uparrow-1}{2J} \right)}.
\end{gather}

We first note that $T$ does not depend on the initial state in $A$ or $B$.  Thus, mixtures of initial states within the $N_\uparrow$ symmetry sector will also exhibit PST, since the period for PST is the same for any initial value.  Therefore, evolution of the mixed state 
\begin{gather}
\label{eq: mixed state swap purity preserving}
    \sum_{x'} p_{x'} |x=x'\rangle \langle x=x'| \leftrightarrow \sum_{x'} p_{x'} |x=N_\uparrow -x'\rangle \langle x=N_\uparrow-x'|
\end{gather}
will swap at period $T$.

The period \eqref{eq: purity preserving period} does, on the other hand, depend on $J$, $\Delta$, and $N_\uparrow$.  This means initial states which are mixtures of these quantum numbers will \textit{not} achieve mixed state transfer, as the different states in the mixture will not swap resonantly.  

For the mixed state transfer in this section, the purity of each subsystem $A$ and $B$ after transfer---for example, for a mixed state of the form \eqref{eq: mixed state swap purity preserving}---is the same as its initial purity (though this purity can change within the swapping period).  We show how to achieve purification of one of the nuclear spin species---by removing the $\Delta$ and $N_\uparrow$ dependence of the swapping period---in the next section.  

\subsubsection{Purification when $\gamma_\text{M},\gamma_\text{J}\neq0$}
\label{Sec: Purification Through Swapping}

We now consider cases where $\rho^A_{\text{initial}}$ and $\rho^B_{\text{initial}}$ can be mixtures of different magnetizations and one species can be a mixture in total spin, i.e. this corresponds to initial states of the form \eqref{eq: initial rho} and \eqref{eq: swap of initial rho}.  This case allows for purification, as one system can be a mixture of different magnetizations while the other subsystem has definite magnetization.  The strategy from the previous section fails to achieve this as it requires the initial state to be in a mixture of different $N_\uparrow$, $\Delta$.  This problem is solved by adding the term ${\cal H}_\text{tune}$ which will allow for the removal of both the $N_\uparrow$ and $\Delta$ dependence in the swapping period.

Using the same basis \eqref{eq: H in pure basis}, note that 
\begin{gather}
    (J^A_z + J^B_z )|x\rangle = (-2J - \Delta + N_\uparrow) |x\rangle, \\
    (\vec{J}^B)^{2} |x \rangle = \left[J(J+1) + (2J+1)\Delta + \Delta^2 \right] |x \rangle .  
\end{gather}
Both of the above are diagonal in $|x\rangle$ and independent of $x$.  This means that our system Hamiltonian in the $|x \rangle$ basis will be of the same form as \eqref{eq: H in pure basis}, except with
\begin{gather}
\label{Eq: Rescaled Cx}
    \gamma_\text{int} C_x \rightarrow J\gamma_\text{int}W(x,N_\uparrow,\Delta) \sqrt{(x+1)(N_\uparrow-x)}
\end{gather}
where
\begin{equation}
\begin{gathered}
\label{eq: W definition}
    W(x,N_\uparrow,\Delta) = \left[ 1 + J\frac{\gamma_M}{\gamma_\text{int}} \left(-2 - \frac{\Delta}{J} + \frac{N_\uparrow}{J} \right) \right. \\
    \left. + J^2\frac{\gamma_J}{\gamma_\text{int}} \left(\frac{J(J+1)}{J^2}+\frac{(2J+1)}{J} \frac{\Delta}{J} + \frac{\Delta^2}{J^2} \right) \right]\\
    \times\sqrt{\left(2-\frac{x}{J}\right) \left(2 + 2 \frac{\Delta}{J} -\frac{N_\uparrow}{J}+\frac{x}{J}+\frac{1}{J}\right)}.
\end{gathered}
\end{equation}

We now go to the limits $\frac{N_\uparrow}{J},\frac{\Delta}{J} \ll 1$.  Here, the $x$ dependence of $W(x,N_\uparrow,\Delta)$ will go to zero up to first order, thus $W$ becomes a constant which only acts to rescale $\gamma_\text{int}$ and (similarly to the previous section) the swapping period will become $T = \frac{\pi}{2\gamma_\text{int} J W(N_\uparrow,\Delta)}$.  Our goal now is to choose $\gamma_M$ and $\gamma_J$ such that the $\frac{N_\uparrow}{J}$ and $\frac{\Delta}{J}$ dependence of $W$ also goes to zero up to first order.  

We find (details in Appendix \ref{Appendix: derivation of gamma_M and gamma_J}) that the required values of $\frac{\gamma_M}{\gamma_\text{int}}$ and $\frac{\gamma_J}{\gamma_\text{int}}$ are 
\begin{gather}
\label{eq: gamma_M and gamma_J}
    \frac{\gamma_M}{\gamma_\text{int}} = \frac{2J+1}{2J (7J+4)} , \;\; \frac{\gamma_J}{\gamma_\text{int}} = \frac{-1}{J (7J+4)}.
\end{gather}
With these parameter values, we find that
\begin{gather}
    T = \frac{\pi}{2 \gamma_\text{int} J} \frac{7J^2 + 4J}{8J^2 + 6J+1}.
    \label{eq: swapping period with Htune}
\end{gather}

\begin{figure}
    \centering
    \includegraphics[width=0.99\linewidth]{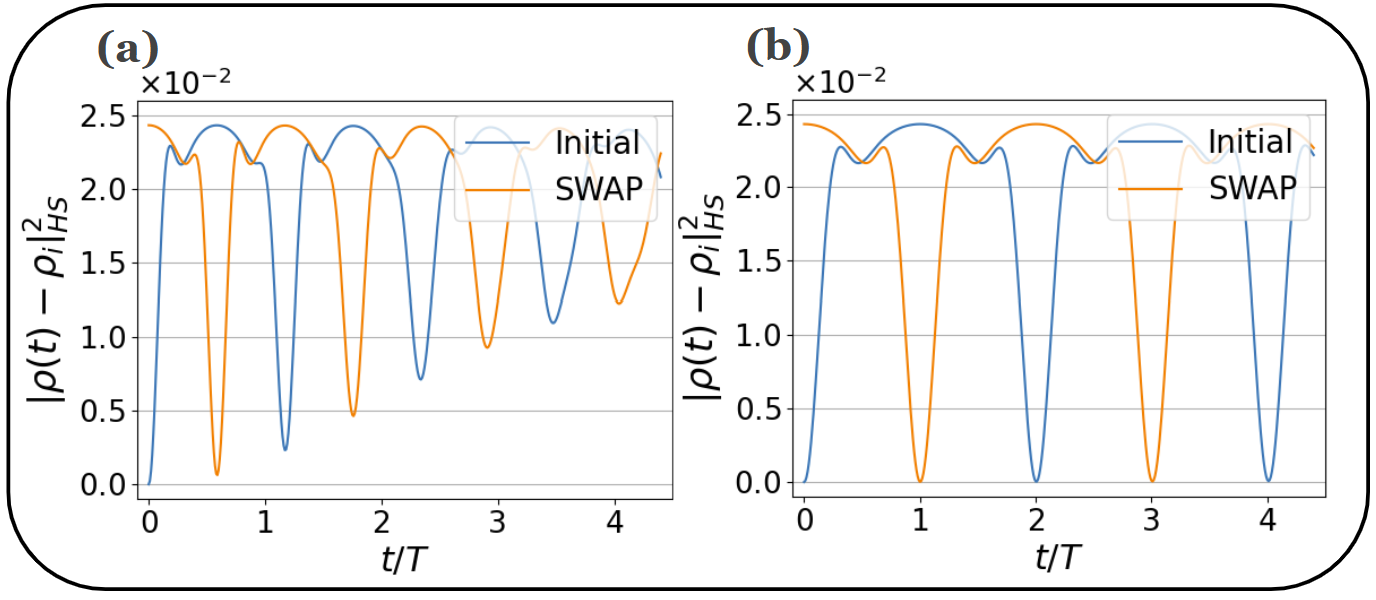}
    \caption{Mixed State Correction:  Initial state is \eqref{eq: initial rho} with parameters given in Sec. \ref{Sec: Purification Through Swapping}.  The system evolves under  (a) only ${\cal H}_\text{int}$ (b) ${\cal H}_\text{int} +{\cal H}_\text{tune}$.  The addition $\gamma_M,\gamma_J \neq 0$ is thus successful in achieving swapping for mixed states.}
    \label{fig:mix of J and M correction}
\end{figure}

Thus, through the introduction of ${\cal H}_\text{tune}$ \eqref{eq: Htune}, we have removed the dependence of $N_\uparrow$ and $\Delta$ for the swapping period and now initial states which are mixtures of these quantum numbers will exhibit mixed state transfer (i.e. the qudit state of the two subsystems will be swapped).  Figure \ref{fig: central spin model and purity pumping} exhibits this swapping for the initial state \eqref{eq: initial rho} with system parameters $J=50$, $\Delta_\text{max}=4$, and $N_{\uparrow,\text{max}}=8$.  In Fig. \ref{fig:mix of J and M correction} we compare evolution of this state with and without the addition of ${\cal H}_\text{tune}$.

We also note here that the desired values for the system parameters exhibit different orders of magnitude in $J$,  $\frac{\gamma_M}{\gamma_\text{int}} = O(\frac{1}{J})$ and $\frac{\gamma_J}{\gamma_\text{int}} = O(\frac{1}{J^2})$.  This is consistent with the fact that, experimentally, $\gamma_M$ will be larger than $\gamma_J$, as higher degree monomials of total spin component will have lower effective coupling in the Schrieffer-Wolff-type transformation used to generate the system Hamiltonian terms and the $\gamma_M$ term of ${\cal H}_\text{tune}$ is a degree $3$ monomial in total spin component operators while the $\gamma_J$ term is degree $4$.   

\section{Robustness to Decoherence and Errors}
\label{Sec: Decoherence}
We now numerically test the robustness of the swapping procedure in the partially polarized limit (Sec. \ref{Sec: Purification Through Swapping}) to decoherence and errors.  In particular, we consider an error $\varepsilon_M$ in the tuning of $\gamma_M(\varepsilon_M)=(1+\varepsilon_M)\gamma_M$ to the desired value found in \eqref{eq: gamma_M and gamma_J} [and similarly for $\gamma_J(\varepsilon_J)$] and we introduce four kinds of decoherence which all act (without loss of generality) on nuclear spin species $A$---collective dephasing, collective decay, local dephasing, and local decay with rates $\gamma_z$, $\gamma_-$, $\kappa_z$, and $\kappa_-$ respectively. Thus,  the master equation for evolution becomes
\begin{equation}
\label{eq: full master equation}
\begin{gathered}
    \dot{\rho} = -\ii  [{\cal H}, \rho] + \gamma_z{\cal D}(J^A_z)[\rho] + \gamma_-{\cal D}(J^A_-)[\rho] \\
    + \kappa_z \sum_i{\cal D}(Z_i^{A})[\rho] + \kappa_- \sum_i{\cal D}(\sigma_{-,i}^{A})[\rho]
\end{gathered}
\end{equation}
where ${\cal D}(L)[\rho]$ is the Lindblad dissipator
\begin{gather}
    {\cal D}(L)[\rho] = L \rho L^\dagger - \frac{1}{2} \{L^\dagger L,\rho\}
\end{gather}
and $Z_i^{A},\sigma_{-,i}^{A}$ the Pauli $z$ and lowering operator for nuclear spin $i$ in $A$ respectively.  The collective dephasing and decay represents coupling to degrees of freedom with long wavelengths (e.g. low frequency magnetic fields), while local dephasing and decay is introduced to model all other short-range perturbations (e.g. local phonons).

\begin{figure}
    \centering
    \includegraphics[width=0.99\linewidth]{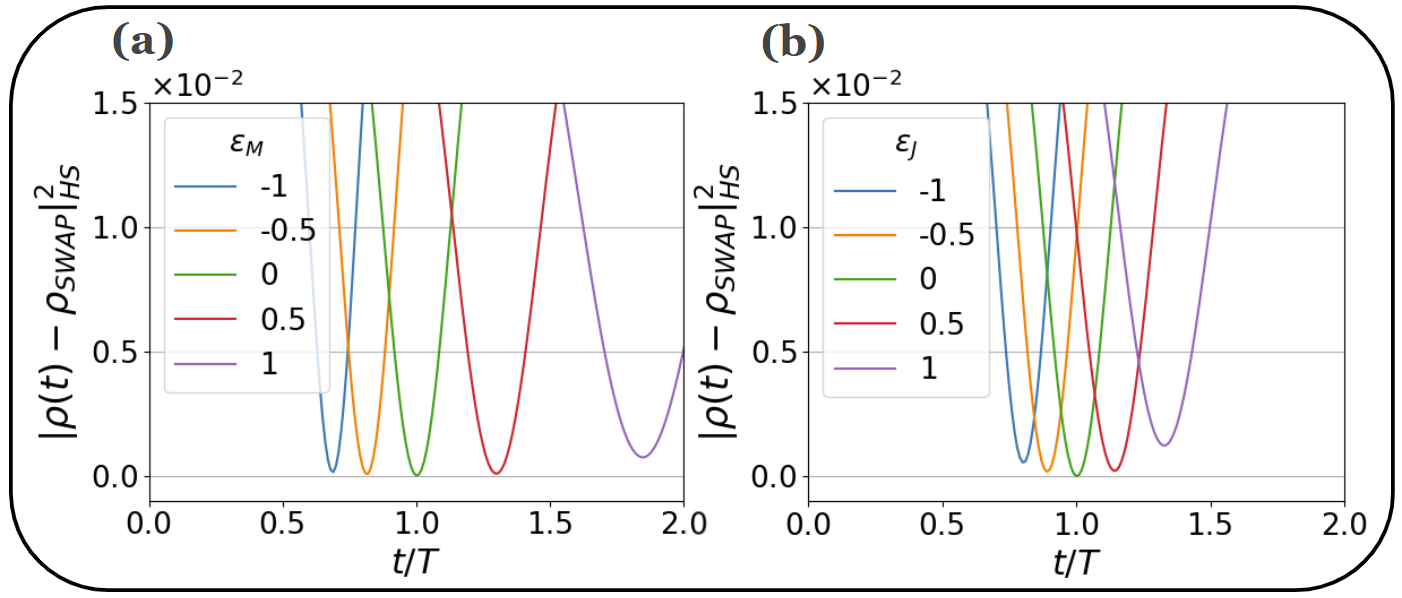}
    \caption{Error in tuning parameters from calculated values \eqref{eq: gamma_M and gamma_J}.  (a) $\gamma_M(\varepsilon_M)=(1+\varepsilon_M)\gamma_M$ (b) $\gamma_J(\varepsilon_J)=(1+\varepsilon_J)\gamma_J$.  Note, here, $\varepsilon_M = -1$ corresponds to $\gamma_M=0$ (and similarly for $\varepsilon_J$).}
    \label{fig:decay and vary gamma_f}
\end{figure}

In Fig. \ref{fig:decay and vary gamma_f} we show the effect of the errors $\varepsilon_M$ and $\varepsilon_J$.  Note, the $\varepsilon_M = -1$ term corresponds to $\gamma_M=0$ and similarly for $\gamma_J$.  In both cases, the accuracy of the swap is maximized at $\varepsilon_M=\varepsilon_J=0$, consistent with Sec. \ref{Sec: Purification Through Swapping}.  For extremely large $\frac{J}{N_\uparrow}$ and $\frac{J}{\Delta}$, however, this correction may be small compared to other sources of error such as decoherence.  In such a context, adding ${\cal H}_\text{tune}$ is not worth the improvement to swapping fidelity and ${\cal H}_\text{int}$ alone is sufficient.  

\begin{figure}
    \centering
    \includegraphics[width=\linewidth]{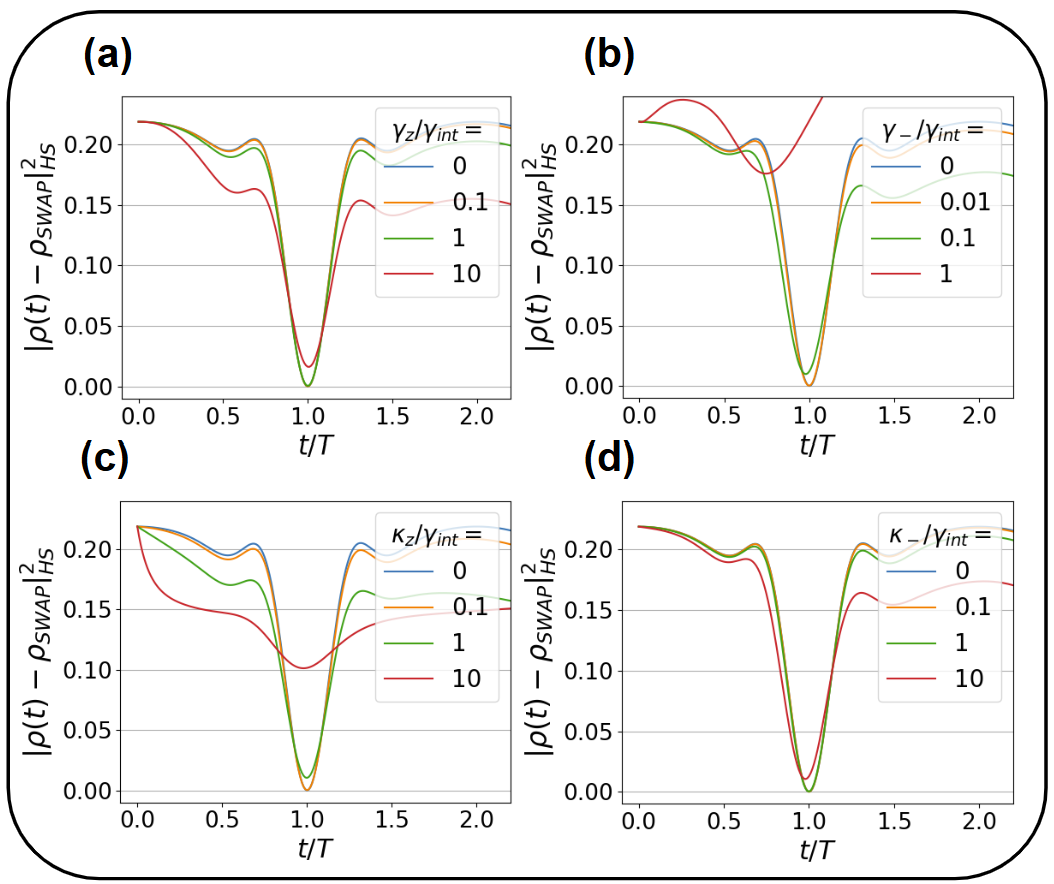}
    \caption{Robustness of Swapping for initial state \eqref{eq: initial rho} with $N_{\uparrow,\text{max}}=8$, $J=50$, $N^A=100$, and $\Delta_\text{max}=0$ and  (a) collective dephasing (b) collective decay (c) local dephasing, and (d) local decay.}
    \label{fig:Decoherence}
\end{figure}

For decoherence, we use order of magnitude heuristic methods to argue (below) that the four forms of decoherence only become relevant to the swap procedure when
\begin{subequations}
\begin{equation}
    \frac{\gamma_-}{\gamma_\text{int}} \geq O(1),
\end{equation}
\begin{equation}
    \frac{\gamma_z}{\gamma_\text{int}} \geq O\left(\frac{J}{N_{\uparrow,\text{max}}} \right),
\end{equation}
\begin{equation}
    \frac{\kappa_z}{\gamma_\text{int}} \geq O\left( \frac{J^2}{N^A} \right),
\end{equation}
\begin{equation}
    \frac{\kappa_-}{\gamma_\text{int}} \geq O\left(\frac{N_{\uparrow,\text{max}}J}{\frac{N^A}{2}-J+N_{\uparrow,\text{max}}} \right),
\end{equation}
\end{subequations}
respectively.  These are consistent with numerics shown in Fig. \ref{fig:Decoherence}, i.e. where state \eqref{eq: initial rho} with system parameters $N_{\uparrow,\text{max}}=8$, $\Delta_\text{max}=0$, $J=50$, and $N^A=100$ is swapped while being exposed to decoherence.

Note, the swapping procedure is especially fragile to the local terms when $\frac{N^A}{2} \gg J$.  This is because this limit implies there are many spin-ups in the system that are trapped in singlets and do not evolve under collective dynamics.  These spin-ups which are `hidden' to the collective dynamics are, however, relevant for the local decoherence and, if there are enough of them, it quickly becomes catastrophic for the swapping procedure.  We find, on the other hand, if $\frac{N^A}{2 J^A}=O(1)$, then the swapping procedure is more robust to the local errors which become relevant when $\frac{\kappa_z}{\gamma_\text{int}},\frac{\kappa_-}{\gamma_\text{int}} \geq O\left(J \right)$.  In this case, collective decay is most fragile, while the other three forms are more robust.

To make these order of magnitude estimates for the scale of the effect of each form of decoherence, it is helpful to consider the 1-norm, defined for any matrix $W$ to be
\begin{gather}
    ||W||_1 \coloneqq \text{max}_j \sum_i |W_{ij}|,
\end{gather}
of each Lindblad super-operator in \eqref{eq: full master equation}.  Since the swapping period is short, $O\left(\frac{1}{J} \right)$, we expect that the decoherence can only significantly alter the fidelity of the swapping procedure if the 1-norm of the superoperator for the decoherence is the same order of magnitude as the superoperator for Hamiltonian evolution.

With this in mind, consider the Hamiltonian term $-\ii [{\cal H}, \rho]$.  From \eqref{eq: H in pure basis}, it may be seen that $||{\cal H}||_1 = O(\gamma_\text{int}J N_\uparrow)$ (since $x$ is at most $O(N_\uparrow)$).  This implies the super-operator (written in the doubled space) $||-\ii ({\cal H}\otimes I - I \otimes {\cal H})||_1 = O(\gamma_\text{int}J N_\uparrow)$.  We may now compare this to the norms of the four forms of decoherence to find an order of magnitude estimate for when the decoherence becomes relevant.     

First, let us consider collective decay.  Note that, similar to the Hamiltonian term, it is also constructed out of biproducts of $J^{\cal M}_\pm$ operators, thus we have $||\gamma_- {\cal D}(J^A_-)||_1 = O(\gamma_- J N_\uparrow)$.  Therefore, collective decay will become relevant to the dynamics when $\frac{\gamma_-}{\gamma_\text{int}} \geq O(1)$.  This fragility is reflected in Fig. \ref{fig:Decoherence}b.

Second, we consider collective dephasing.  Here, the dissipator is constructed out of biproducts of $J_z^A$.  For a generic density matrix in $A$, $\rho^A = |M^A \rangle |M'^A \rangle = |-J^A+x \rangle |-J^A+y \rangle$ where $\frac{x}{J^A},\frac{y}{J^A} \ll 1$ in the partially polarized limit, we have that the collective dephasing dissipator acts as
\begin{equation}
\label{eq: action of collective dephasing}
\begin{gathered}
    {\cal D}(J_z^A)\rho = \left[M^A M'^A - \frac{1}{2}((M^A)^2 + (M'^A)^2)\right]\rho \\= \left[xy - \frac{1}{2}(x^2+y^2) \right]\rho.
\end{gathered}
\end{equation}
Thus we have that $||\gamma_z {\cal D}(J_z^A)||_1 = O(\gamma_z N_\uparrow^2)$.  This implies that the swapping dynamics will be much more robust against collective dephasing, which will only become relevant when $\frac{\gamma_z}{\gamma_\text{int}} = O(\frac{J}{N_\uparrow})$.  This is verified numerically in Fig. \ref{fig:Decoherence}a.  

For local dephasing, note that
\begin{gather}
    ||\sum_{i\in A} {\cal D}(Z_{i})||_1 =||\sum_{i\in A} (Z_i \otimes Z_i - I)||_1 \\
    = ||\sum_{i \in A} Z_i \otimes Z_i||_1 - N^A,
\end{gather}
therefore it is sufficient to calculate $||\sum_{i\in A} Z_i \otimes Z_i||_1$.  Each $||Z_i \otimes Z_i||_1$ is maximized by $1-O(\frac{N_\uparrow}{J})$ in the $J^A,M^A,\xi^A$ basis, so we have that
\begin{gather}
    ||\sum_{i\in A} {\cal D}(Z_i)||_1 
    = ||\sum_{i\in A} Z_i \otimes Z_i||_1 - N^A \\
    \leq N^A \text{max}_{i\in A}|| Z_i \otimes Z_i||_1 - N^A = O(N^A \frac{N_\uparrow}{J}), 
\end{gather}
where in the second line we have used the triangle inequality.  In Appendix \ref{Appendix: mixed computational details}, we show that this bound is tight.  Thus, local dephasing will become relevant when $\frac{\kappa_z}{\gamma_\text{int}} \geq O( \frac{J^2}{N^A})$. The case $J = \frac{N^A}{2}$ is shown in Figure \ref{fig:Decoherence}c.

For local decay, note that ${\cal D}(\sigma_{-,i}^{A})|1 \rangle_{A,i} |1\rangle_{A,i} = |0\rangle_{A,i} | 0\rangle_{A,i} - |1 \rangle_{A,i} | 1\rangle_{A,i}$ and $0$ for any other state at site $i\in A$.  This means that $||\kappa_- \sum_{i\in A} {\cal D}(\sigma_{-,i})||_1 = O(\kappa_-(\frac{N^A}{2}-J+N_{\uparrow,\text{max}}))$, where $\frac{N^A}{2}-J+N_{\uparrow,\text{max}}$ is an upper bound for the total number of spin-ups which could be in the $A$ subsystem.  This implies it will begin to significantly affect the accuracy of the swapping procedure when $\frac{\kappa_-}{\gamma_\text{int}} \geq O\left(\frac{N_{\uparrow,\text{max}} J}{\frac{N^A}{2}-J+N_{\uparrow,\text{max}}} \right)$.  Note, again as expected for local decoherence, the swapping protocol is fragile if $N^A \gg J$.  However, if $J$ is near $\frac{N^A}{2}$, then the decoherence only becomes relevant when $\frac{\kappa_-}{\gamma_\text{int}} \geq O\left(J \right)$.  This case is shown in Fig. \ref{fig:Decoherence}d.

As an aside, we also note that swapping may in fact be used to protect against decoherence.  This is because, under the swapping dynamics, the state of the system in $A$ spends part of its time in subsystem $B$ where there is no decoherence.  This is verified numerically in Appendix \ref{Appendix: Swapping Protects}.

\section{Conclusion and Discussion}
\label{Sec: Conclusion}
We have shown that it is possible to engineer a Hamiltonian between the nuclear spins of a quantum dot (or a more general central spin system) which swaps the quantum state of two nuclear spin species.  In particular, when the two nuclear spin species are partially polarized and have similar total spin, arbitrary mixed states may be swapped between the two subsystems.  We did a preliminary analysis of decoherence in the system and found that the swap (when total spin is near its maximal value) is most fragile to collective decay, while it is robust to collective dephasing, local dephasing, and local decay.  On the other hand, if $2J$ is much less than the number of sites for each nuclear spin species, then much weaker local decoherence can still be prohibitive to achieving the swap protocol.  For realistic implementation of the swap protocol, the swapping period must occur on a timescale which is short compared with the decoherence but long compared with timescales which may be resolved experimentally.  We note that, for example, a $\pi$-gate with the Hamiltonian \eqref{eq: Hint} was found to be implementable across a variety of candidate material platforms in \cite{Zaporski2023preparesinglet}.  This is suggestive that the state transfer protocol may be realizable in the near-term.

While in this work we have focused on the partially polarized and $J^A \approx J^B$ limits (as this is experimentally relevant in many cases, for example, swapping or purifying states stored in the nuclear spins as a quantum register \cite{Appel2025quantumregister}), it would be interesting to extend the procedure beyond such constraints.  The connection to perfect state transfer provides a general method to pursue this aim.  Namely, first add new Hamiltonian terms which will make the full Hamiltonian commute with $P$ as well as provide additional tunable parameters in the system.  Then solve the inverse eigenvalue problem (discussed in Sec. \ref{Sec: PST Review}) to find values of these parameters which ensure odd commensurability of the Hamiltonian eigenvalues, thereby achieving state transfer.  It may additionally be beneficial to instead pursue constructing a Hamiltonian which satisfies the conditions for \textit{almost} perfect state transfer \cite{Vinet2012APST} (by which we mean state transfer is achieved inexactly but within arbitrary precision) which are weaker conditions than those for perfect state transfer discussed in this work.       

It may also be possible to leverage the connection to perfect state transfer in order to realize other quantum gates between the nuclear spin species.  This may be accomplished by choosing $P$ to be the desired gate (instead of the $AB$ swap) and engineering an odd commensurate Hamiltonian which commutes with $P$.  

\section{Acknowledgements}
We are grateful to Dorian Gangloff for stimulating discussions and feedback. This work was supported by a Simons Investigator Award (Grant No. 511029) and the Engineering and Physical Sciences Research Council [grant numbers EP/V062654/1 and EP/Y01510X/1].

\bibliography{main}

\onecolumngrid
\appendix

\section{Weighted Hilbert-Schmidt Norm}
\label{Appendix: HS norm}
We here describe why the regular Hilbert-Schmidt norm can be misleading when describing the distance between the state of the system and the initial or swapped configurations, and why a weighted Hilbert-Schmidt norm is more suitable.  The Hilbert-Schmidt norm of a matrix $W$ is defined as
\begin{gather}
    |W|_\text{HS}^2 = \sum_{ij} |W_{ij}|^2 = \Tr\{W W^\dagger\}.
\end{gather}
For a permutation symmetric density matrix of the form \eqref{eq: permutation symmetric density matrix}, we have that
\begin{gather}
    |\rho|_\text{HS}^2 = \sum_{J,M,M'} \sum_\xi\left|\frac{\rho_{J,M,M'}}{d_J} \right|^2 = \sum_{J,M,M'} \frac{1}{d_J} |\rho_{J,M,M'}|^2.
    \label{eq: regular HS is bad}
\end{gather}
Note, on the other hand, that any observable $\cal O$ of interest for the system will act identically for different $\xi$ and be of the form
\begin{gather}
    \Tr\{\rho{\cal O}\} = \sum_{\Tilde{J},\Tilde{M},\Tilde{\xi}} \langle \Tilde{J},\Tilde{M},\Tilde{\xi}| \rho {\cal O} |\Tilde{J},\Tilde{M},\Tilde{\xi} \rangle = \sum_{\Tilde{J},\Tilde{M}} \langle \Tilde{J},\Tilde{M}|\sum_{J,M,M'} \rho_{J,M,M'} |J,M \rangle \langle J,M'| {\cal O} \left( \frac{1}{d_J}\sum_{\xi,\Tilde{\xi}} \langle \Tilde{\xi} | \xi \rangle \langle \xi | \Tilde{\xi} \rangle \right) |\Tilde{J},\Tilde{M} \rangle \\
    = \sum_{J,M}  \rho_{J,M,M'} \langle J,M'| {\cal O} |J,M\rangle
    \label{eq: Observables of rho}
\end{gather}
In contrast to the Hilbert-Schmidt norm, observables are not weighted by a factor $\frac{1}{d_J}$.  This implies that, compared with observables, the Hilbert-Schmidt norm will be biased towards $J$ where $d_J$ is small (such as for maximal $J=\frac{N}{2}$ where $d_J=1$).  Of course, if $\rho(T)=\rho_\text{SWAP}$ exactly at some $T$, then $|\rho(T)-\rho_\text{SWAP}|_\text{HS}^2=0$ and $\Tr\{\rho(T) {\cal O}\} = \Tr\{\rho_\text{SWAP} {\cal O}\}$.  However, if there is some small error $\rho(T)-\rho_\text{SWAP} = \Delta \rho$, then the errors $\Delta \rho_{J,M,M'}$ for each of the $J,M,M'$ sectors are weighted equally for any observable \eqref{eq: Observables of rho}.  For the Hilbert-Schmidt norm, on the other hand, errors $\Delta \rho_{J,M,M'}$ are suppressed by a factor $\frac{1}{d_J}$ \eqref{eq: regular HS is bad}.  The crux of the issue is that observables are linear in $\rho$ while the Hilbert-Schmidt norm is non-linear.  This biases the Hilbert-Schmidt norm to sectors where the state of the system is not spread out over a mixture of a large number of $\xi$.  

For assessing the fidelity of the swapping procedure, the regular Hilbert-Schmidt norm is thus a suboptimal choice.  This is particularly true in the case $\frac{N^A}{2} > J^A$ or similarly for $B$.  For example, if for some swapping procedure the weight of the errors $\Delta_{J^A-10,M^A,M'^A} = \Delta_{J^A+10,M^A,M'^A}$, the Hilbert-Schmidt norm weights the $J^A+10$ much more than the error in the $J^A-10$ sector, by a factor of $\frac{d_{J^A-10}}{d_{J^A+10}}$.  A naive interpretation of the norm would thus seem to imply that errors are concentrated in sectors with larger $J^A$; since the swapping period is order $J^A$, this makes the swapping period seem shorter than it is.

To fix this problem, it is helpful to consider a weighted Hilbert-Schmidt norm.  Namely,
\begin{gather}
    |\rho|_\text{HS}^2 \rightarrow \sum_{J,M,M'}  |\rho_{J,M,M'}|^2.
\end{gather}
This removes the $\frac{1}{d_J}$ weighting caused by the non-linearity, makes the norm more representative of errors which will actually be measured by observables, and also retains all the usual properties of matrix norms so long as the $\xi$ are left invariant under the dynamics (which they are for all cases considered in this work).

We also note that the 1-norm, used in Sec. \ref{Sec: Decoherence}, is linear in $\rho$ and therefore does not need to be weighted.  

\section{Necessary and Sufficient Conditions for Magnetization Inversion}
\label{Appendix: Necessary and Sufficient Conditions}
We here show that the conditions given in Section \ref{Sec: PST Review} are necessary and sufficient for magnetization inversion as well as show that $T$ in the odd commensurate eigenvalue condition corresponds to the swapping time.  These results mostly follow similar proof given for perfect state transfer in \cite{Kay2010PSTReview}.

\underline{Necessary:} 
To show $H_\text{MI}$ must be mirror symmetric, we first expand the initial and final state in the basis of eigenvectors
\begin{gather}
    |M\rangle = \sum_a \alpha_a |\lambda_a \rangle \\
    |-M \rangle = \sum_a \beta_a | \lambda_a\rangle.    
\end{gather}
Since we have that
\begin{gather*}
    \ee^{-\ii  H_\text{MI} T}|M\rangle = \ee^{-\ii  \phi}|-M \rangle,
\end{gather*}
this implies that
\begin{gather}
    \ee^{-\ii  \lambda_a T} \alpha_a = \ee^{-\ii  \phi} \beta_a \; \forall a \nonumber \\
    \implies |\alpha_a|^2 = |\beta_a|^2.
\end{gather}
We must therefore have that 
\begin{gather}
    \langle J| H_\text{MI}^m | J \rangle = \sum_a \lambda_a^m |\alpha_a|^2 = \langle -J|H_\text{MI}^m |-J\rangle.
\end{gather}
When $m=1$, this implies that $V_{-J} = V_{J}$.  For $m=2$ this implies $V_{-J}^2 + C_{-J}^2 = V_J^2 + C_{J-1}^2 \implies  C_{-J}^2 = C_{J-1}^2$.  By induction this gives that all coefficients satisfy the reflection symmetry $V_M = V_{-M}$ and $C_{-M}^2=C_{M-1}^2$ and therefore $[H_\text{MI},P]=0$.

For the odd commensurate eigenvalue condition, since $[H_\text{MI},P]=0$, we may simultaneously diagonalize to find $P|\lambda_a\rangle = (-1)^{N_a}|\lambda_a \rangle$ where $N_a$ is an $a$ dependent integer which partitions the eigenvalues into those symmetric under $P$ ($N_a$ even) versus anti-symmetric under $P$ ($N_a$ odd).  As $H_\text{MI}$ \eqref{eq: H for pure magnetization inversion} is a Jacobi matrix, the Sturm Oscillation theorem may be used to show that if $|\lambda_a\rangle$ is symmetric under $P$, then $|\lambda_{a \pm 1}\rangle$ is anti-symmetric under $P$ (and vice versa) \cite{Kay2010PSTReview,Gladwell2005InverseProblems}.  Thus, the difference of neighboring $N_a$ must be odd, i.e. $N_{a+1}-N_a = 2 m_a+1$ for some integer $m_a$.  We therefore have that
\begin{gather}
    \ee^{-\ii  H_\text{MI} T} = \ee^{\ii \phi} P \implies \ee^{-\ii  \lambda_a T} = \ee^{\ii \phi} (-1)^{N_a} \\
    \implies \lambda_aT = \phi + \pi N_a + 2 \pi c_a \implies (\lambda_{a+1}-\lambda_a)T = \pi [N_{a+1}-N_a + 2(c_{a+1}-c_a)].
\end{gather}
In the above, $N_{a+1}-N_a$ is odd and $2(c_{a+1}-c_a)$ is even, thus $[N_{a+1}-N_a + 2(c_{a+1}-c_a)]$ must be odd, giving exactly the odd commensurability condition.

\underline{Sufficient:} From the odd commensurability condition, we have that
\begin{gather}
    \lambda_a = \sum_{b=0}^{a-1}(2 m_b +1) \frac{\pi}{T} + \lambda_0 \equiv N_a \frac{\pi}{T}+ \lambda_0
\end{gather} 
where we have defined $N_a= \sum_{b=0}^{a-1}(2 m_b +1)$.  Note that $N_a$ is odd for odd $a$ and even for even $a$.  

We now write
\begin{gather}
    \ee^{-\ii  H_\text{MI} T}|M\rangle = \sum_a \alpha_a \ee^{-\ii  \lambda_a T}|\lambda_a \rangle = \ee^{-\ii  \lambda_0 T} \sum_a \alpha_a (-1)^{N_a} |\lambda_a \rangle = \ee^{\ii \phi} P |M \rangle,
\end{gather}
where in the last line we have defined $\phi = - \lambda_0 T$ and used $P |\lambda_a \rangle = (-1)^{N_a} |\lambda_a \rangle$ (which, as in the proof for the necessary conditions, comes from mirror symmetry combined with the Sturm oscillation theorem).  This also shows that the swapping period is given precisely by $T$.    

\section{Mixed State Computation Details}
\label{Appendix: mixed computational details}
In this Appendix, we give calculation details for all results for mixed states.  As noted in section \ref{Sec: Permutation Symmetric Mixed States}, both $\rho^A$ and $\rho^B$ must be of the form \eqref{eq: permutation symmetric density matrix}.  We give how the system parameters $J^A,M^A,M'^A,J^B,M^B,M'^B$ transform under the full system master equation \eqref{eq: full master equation}.  This allows us to describe the mixed state transfer problem, of the form \eqref{eq: mixed PST}, in this basis and draw connections to the pure state solution found in Section \ref{Sec: Mixed State Transfer}.  This basis also allows for the efficient simulation of decoherence used to generate Fig. \ref{fig:Decoherence}, we here provide details.  The 1-norm estimates from Sec. \ref{Sec: Decoherence} for each term in the master equation of the system are also verified.

As a starting point, we rewrite \eqref{eq: permutation symmetric density matrix} for subsystem $A$ and $B$ in the doubled space to characterize the density matrix of the full system
\begin{gather}
    \rho = \sum_{\substack{J^A,M^A,M'^A\\J^B,M^B,M'^B}} \rho_{J^A,M^A,M'^A} \rho_{J^B,M^B,M'^B}\overline{|J^A,M^A \rangle | J^A,M'^A\rangle} \;  \overline{|J^B,M^B \rangle | J^B,M'^B\rangle}.
    \label{eq: unsimplified rho rho}
\end{gather}
We now simplify \eqref{eq: unsimplified rho rho} using the symmetries and constraints of our dynamics \eqref{eq: full master equation}.  First, subsystem $B$ is only acted on by collective terms which preserve $J^B$.  Thus, the initial mixture of $J^B$ will be conserved and we need only keep track of the magnetization; as such, we will neglect writing the sum over $J^B$ and write the state $|J^B,M^B \rangle \rightarrow|M^B \rangle$.  As we are working in the partially polarized limit, it is helpful (similarly to the basis \eqref{eq: pure state transfer basis}) to rewrite the magnetizations in terms of fluctuations from the fully polarized state, i.e. for example $M^A = -J^A + x$ where $x$ is the change from the fully polarized state.  It is also helpful to write the state in a basis which leverages the fact that ${\cal H}$ preserves total magnetization.  All together, we then have that

\begin{gather}
    \rho = \sum_{J^A,N_\uparrow,x,y} \rho_{J^A,N_\uparrow,x,y} \overline{|J^A,-J^A+x \rangle | J^A,-J^A+N_\uparrow-y\rangle} \;  \overline{|-J^B+N_\uparrow-x \rangle |-J^B+y\rangle}.
    \label{eq: simplified rho rho}
\end{gather}
The indices take the values $J^A \in \{0,1,\ldots,\frac{N^A }{2}\}$, $N_\uparrow \in \{0,1,\ldots,2J^A \}$, $x \in \{0,1,\ldots,N_\uparrow \}$, and $y \in \{0,1,\ldots,N_\uparrow \}$.  We will find, however, that under the master equation \eqref{eq: full master equation}, evolution is restricted to an even smaller set $J^A \in \{J^A_{\text{initial}}-N_{\uparrow,\text{initial}},\ldots,\frac{N^A}{2}\}$ and $N_\uparrow \in \{0,\ldots,(\frac{N^A}{2}-J^A_{\text{initial}}) + N_{\uparrow,\text{initial}} \}$.  Furthermore, the $N_\uparrow \ll J^A$ condition will always remain satisfied.

\subsection{Coherent Evolution}
If there is no decoherence, then $J^A$ and $N_\uparrow$ are also conserved.  Thus, the space of density matrices becomes
\begin{gather}
    \rho = \sum_{x,y} \rho_{x,y} \overline{|-J^A+x \rangle |-J^A+N_\uparrow-y\rangle} \;  \overline{|-J^B+N_\uparrow-x \rangle |-J^B+y\rangle}.
    \label{eq: rho rho without decoherence}
\end{gather}
We first consider evolution under ${\cal H}_\text{int}$.  The Hamiltonian evolution in the doubled space becomes
\begin{gather}
    -\ii[{\cal H}_\text{int},\rho] \rightarrow -\ii \gamma_\text{int} \left[ \left(J_+^A J_-^B + J_-^A J_+^B \right) \otimes I - I \otimes  \left(J_+^A J^B_- + J_-^A J_+^B \right)  \right] \rho.
\end{gather}
Using $C_x$ from \eqref{eq: Cx from H in pure basis}, we therefore have that evolution under ${\cal H}_\text{int}$ evolves \eqref{eq: rho rho without decoherence} in the following way
\begin{subequations}
\begin{equation}
    x \rightarrow x+1: \;\; -\ii \gamma_\text{int} C_x,
\end{equation}
\begin{equation}
    x+1 \rightarrow x: \;\; -\ii \gamma_\text{int} C_x,
\end{equation}
\begin{equation}
    y \rightarrow y+1: \;\; \ii \gamma_\text{int} C_y,
\end{equation}
\begin{equation}
    y+1 \rightarrow y: \;\; \ii \gamma_\text{int} C_y.
\end{equation}
\end{subequations}
Note that this is just two independent copies of the pure state transfer \eqref{eq: H in pure basis}, one for the left Hilbert space ($x$) and one for the right Hilbert space ($y$).  Therefore, one just needs to solve the pure state transfer problem---as done in Sec. \ref{Sec: Swap with just Hint}---to find when $x \leftrightarrow N_\uparrow - x$, and then $y \leftrightarrow N_\uparrow - y$ will swap at the same rate.  Adding ${\cal H}_\text{tune}$ simply rescales $\gamma_\text{int}$ in the same way it did for the pure case, i.e. \eqref{Eq: Rescaled Cx}.

\subsection{Decoherence}
We now return to the form \eqref{eq: simplified rho rho} and calculate how $\rho$ is transformed under the superoperators for the four forms of decoherence \eqref{eq: full master equation}.  

First, collective dephasing.  In a similar fashion to \eqref{eq: action of collective dephasing}, we find that 
\begin{equation}
\begin{gathered}
    {\cal D}(J_z^A)\rho = \sum_{J^A,N_\uparrow,x,y}\left[x(N_\uparrow-y)-\frac{1}{2} \left(x^2+(N_\uparrow-y)^2 \right) \right] \rho_{J^A,N_\uparrow,x,y}\\
    \times \overline{|J^A,-J^A+x \rangle | J^A,-J^A+N_\uparrow-y\rangle} \;  \overline{|-J^B+N_\uparrow-x \rangle |-J^B+y\rangle}.
\end{gathered}
\end{equation}
Thus, ${\cal D}(J_z^A)$ acts diagonally on the basis with a factor $\left[x(N_\uparrow-y)-\frac{1}{2} \left(x^2+(N_\uparrow-y)^2 \right) \right]$.  We denote this by\\\\
\underline{Collective Dephasing:}
\begin{gather}
    \text{diag}: \;\; \left[x(N_\uparrow-y)-\frac{1}{2} \left(x^2+(N_\uparrow-y)^2 \right) \right]
\end{gather}

For collective decay, let us take the superoperator
\begin{gather}
    {\cal D}(J_-^A) = J_-^A \otimes J_-^A - \frac{1}{2} \left(J_+^A J_-^A \otimes I + I \otimes J_+^A J_-^A \right)
    \label{eq: dissipator for collective decay}
\end{gather}
and focus on the two terms seperately.  The first term lowers magnetization by 1 (with the normal prefactor $J_- |J,M \rangle = \sqrt{(J+M)(J-M+1)}$) for the left and right Hilbert space of subsystem $A$.  We therefore have $-J^A+x \rightarrow -J^A+x-1$ and $-J^A+N_\uparrow-y \rightarrow -J^A+N_\uparrow-1-y$, or in other words $x \rightarrow x-1$, $N_\uparrow \rightarrow N_\uparrow-1$.  The second term in \eqref{eq: dissipator for collective decay} is diagonal in $J^A,N_\uparrow,x,y$ and gives the factor $-\frac{1}{2}\left(x(2J^A-x+1)+(N_\uparrow-y)(2J^A-N_\uparrow+y+1) \right)$.  In summary, we have \\\\
\underline{Collective Decay:}
\begin{subequations}
\begin{gather}
    \substack{x \rightarrow x-1\\ N_\uparrow \rightarrow N_\uparrow-1}: \;\; \sqrt{x(2J^A-x+1)(N_\uparrow-y)(2J^A-N_\uparrow+y+1)},
\end{gather}  
\begin{gather}
    \text{diag}: \;\;  -\frac{1}{2}\left(x(2J^A-x+1)+(N_\uparrow-y)(2J^A-N_\uparrow+y+1) \right)
\end{gather}
\end{subequations}

Solving for the effect of the local terms is more involved.  The general form for symmetric, local Pauli operations was found in \cite{Chase2008PermutationSymmetric}.  Using this result, we find that\\\\
\underline{Local Dephasing:}
\begin{subequations}
\begin{gather}
    \substack{J^A \rightarrow J^A-1 \\ N_\uparrow \rightarrow N_\uparrow-1 \\ x \rightarrow x-1}:\;\; \frac{2(\frac{N^A}{2}+J^A+1)\sqrt{x(2J^A-x)(N_\uparrow-y)(2J^A-N_\uparrow+y)}}{ J^A (2 J^A+1)}
\end{gather}
\begin{gather}
    \substack{J^A \rightarrow J^A+1 \\ N_\uparrow \rightarrow N_\uparrow+1 \\ x \rightarrow x+1}:\;\; \frac{2(\frac{N^A}{2}-J^A)\sqrt{(x+1)(2J^A-x+1)(N_\uparrow-y+1)(2J^A-N_\uparrow+y+1)}}{(J^A+1) (2 J^A+1)}
\end{gather}
\begin{gather}
    \text{diag}:\;\; \frac{2(\frac{N^A}{2}+1)(-J^A+x)(-J^A+N_\uparrow-y)}{ J^A ( J^A+1)} - N^A
\end{gather}
\end{subequations}
and\\\\
\underline{Local Decay:}
\begin{subequations}
\begin{gather}
    \substack{N_\uparrow \rightarrow N_\uparrow-1 \\ x \rightarrow x-1}:\;\; \frac{(\frac{N^A}{2}+1)\sqrt{x(2J^A-x+1)(N_\uparrow-y)(2J^A-N_\uparrow+y+1)}}{2 J^A (J^A+1)}
\end{gather}
\begin{gather}
    \substack{J^A \rightarrow J^A-1 \\ N_\uparrow \rightarrow N_\uparrow-2 \\ x \rightarrow x-2}:\;\; \frac{(\frac{N^A}{2}+J^A+1)\sqrt{x(x-1)(N_\uparrow-y)(N_\uparrow-y-1)}}{2 J^A (2J^A+1)}
\end{gather}
\begin{gather}
    J^A \rightarrow J^A+1:\;\; \frac{(\frac{N^A}{2}-J^A)\sqrt{(2J^A-x+1)(2J^A-x+2)(2J^A-N_\uparrow+y+1)(2J^A-N_\uparrow+y+2)}}{2 (J^A+1)(2J^A+1)}
\end{gather}
\begin{gather}
    \text{diag}:\;\; - \frac{N^A-2J^A+x+N_\uparrow-y}{2}
\end{gather}
\end{subequations}

Note, in the local terms above, $J^A$ may only be decreased if $N_\uparrow$ is decreased.  This means $J^A$ may never be less than $J^A_{\text{initial}}-N_{\uparrow,\text{initial}}$.  On the other hand, $J^A$ can increase on its own, so can reach its maximum value of $J^A = \frac{N^A}{2}$.  The index $N_\uparrow$ may only increase if $J^A$ is also increased.  This means $N_\uparrow$ may not be increased by more than $\frac{N^A}{2}-J^A_{\text{initial}}$ above its initial value.  Note, this corresponds to the extra spin-ups in the system, discussed Sec. \ref{Sec: Decoherence}, which are trapped in singlets and thereby hidden to the collective dynamics; these spin-ups may, however, be freed from the singlets due to the local decoherence and thus increase $N_\uparrow$ (the number of spin-ups which may be transferred between $A$ and $B$ under the coherent evolution).  These useful bounds on $J^A$ and $N_\uparrow$ allow for efficient numerical simulation when $J^A$ is close to its maximal value.  

The order of magnitude estimates for the 1-norm of each of the superoperators in \eqref{eq: full master equation}, discussed in Sec. \ref{Sec: Decoherence}, may all be verified using their explicit form found in this section.

\section{Derivation of $\gamma_M$ and $\gamma_J$ for swapping}
\label{Appendix: derivation of gamma_M and gamma_J}
We here derive the values of $\gamma_J$ and $\gamma_M$ which remove the $\Delta = J^B - J^A$ and $N_\uparrow$ dependence of the swapping period up to first order in $\frac{N_\uparrow}{J}$ and $\frac{\Delta}{J}$ ($J^A \equiv J$).  As discussed in Sec. \ref{Sec: Purification Through Swapping}, this is achieved by finding values of $\gamma_M$ and $\gamma_J$ where $W(x,N_\uparrow,\Delta)$ \eqref{eq: W definition} becomes independent of $x$, $N_\uparrow$, and $\Delta$ up to first order.  We reproduce $W$ here for convenience   
\begin{equation}
\begin{gathered}
    W(x,N_\uparrow,\Delta) = \left[ 1 + J\frac{\gamma_M}{\gamma_\text{int}} \left(-2 - \frac{\Delta}{J} + \frac{N_\uparrow}{J} \right) + J^2\frac{\gamma_J}{\gamma_\text{int}} \left(\frac{J(J+1)}{J^2}+\frac{(2J+1)}{J} \frac{\Delta}{J} + \frac{\Delta^2}{J^2} \right) \right]\\
    \times\sqrt{\left(2-\frac{x}{J}\right) \left(2 + 2 \frac{\Delta}{J} -\frac{N_\uparrow}{J}+\frac{x}{J}+\frac{1}{J}\right)}.
    \label{eq: W definition in appendix}
\end{gathered}
\end{equation}
Note, from the form of \eqref{eq: W definition in appendix}, we expect that the solution will be of the order $\gamma_M = O(\frac{1}{J})$ and $\gamma_J = O(\frac{1}{J^2})$.

The square root in \eqref{eq: W definition in appendix} may be expanded to find
\begin{gather}
    \sqrt{\left(2-\frac{x}{J}\right) \left(2 + 2 \frac{\Delta}{J} -\frac{N_\uparrow}{J}+\frac{x}{J}+\frac{1}{J}\right)} = 2 + \frac{\Delta}{J} - \frac{1}{2} \frac{N_\uparrow}{J} + \frac{1}{2 J} + \text{higher orders}
\end{gather}
We therefore have that 
\begin{equation}
\begin{gathered}
    W(x,N_\uparrow,\Delta) = \left[ 1 + J\frac{\gamma_M}{\gamma_\text{int}} \left(-2 - \frac{\Delta}{J} + \frac{N_\uparrow}{J} \right) + J^2\frac{\gamma_J}{\gamma_\text{int}} \left(\frac{J(J+1)}{J^2}+\frac{(2J+1)}{J} \frac{\Delta}{J} \right) \right]\left(2 + \frac{\Delta}{J} - \frac{1}{2} \frac{N_\uparrow}{J} + \frac{1}{2 J} \right) \\
    + \text{higher orders}
\end{gathered}
\end{equation}
\begin{equation}
\begin{aligned}
    = &\left(2 + \frac{1}{2J} \right) \left(1 - 2J \frac{\gamma_M}{\gamma_\text{int}} + \frac{\gamma_J}{\gamma_\text{int}} J (J+1) \right) \\
    &+ \frac{\Delta}{J} \left[1 - 4J \frac{\gamma_M}{\gamma_\text{int}} + J (3J+2) \frac{\gamma_J}{\gamma_\text{int}}  \right]\\
    &+ \frac{N_\uparrow}{J} \left[-\frac{1}{2} + 3 J \frac{\gamma_M}{\gamma_\text{int}}  - \frac{1}{2} J (J+1) \frac{\gamma_J}{\gamma_\text{int}} \right]\\
    &+ \text{higher orders}
\end{aligned}
\end{equation}
To get rid of the factors of $\frac{\Delta}{J}$ and $\frac{N_\uparrow}{J}$ we must solve the system of equations
\begin{gather}
    1 - 4J \frac{\gamma_M}{\gamma_\text{int}} + J (3J+2) \frac{\gamma_J}{\gamma_\text{int}} = 0 \\
    -\frac{1}{2} + 3 J \frac{\gamma_M}{\gamma_\text{int}}  - \frac{1}{2} J (J+1) \frac{\gamma_J}{\gamma_\text{int}} = 0
\end{gather}
The solution is
\begin{gather}
    \frac{\gamma_M}{\gamma_\text{int}} = \frac{2J+1}{2J (7J+4)} , \;\; \frac{\gamma_J}{\gamma_\text{int}} = \frac{-1}{J (7J+4)}.
\end{gather}
Finally, we have 
\begin{gather}
    W(x,N_\uparrow,\Delta) \rightarrow \frac{8 J^2 +6J+1}{7J^2+4J}
\end{gather}
which yields the swapping period given in \eqref{eq: swapping period with Htune}.

\section{Swapping Protects From Decoherence}
\label{Appendix: Swapping Protects}
\begin{figure}[h]
    \centering
    \includegraphics[width=0.4\linewidth]{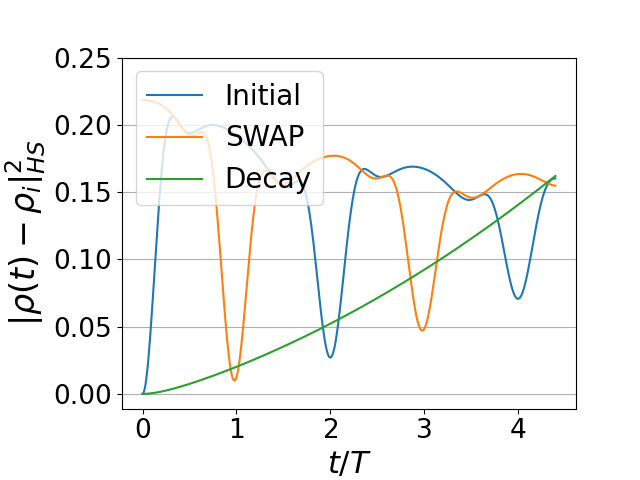}
    \caption{Swapping Protects Against Decoherence: The initial state \eqref{eq: initial rho} evolving under the swapping protocol and collective decay is compared with the state evolving solely under collective decay.  By decay in the legend above, we mean the y-axis is given by $|\rho(t)-\rho_\text{initial}|_\text{HS}^2$ where $\rho(t)$ is the evolution under collective decay alone.}
    \label{fig:swap protects decay}
\end{figure}

As discussed at the end of Sec. \ref{Sec: Decoherence}, swapping can help preserve a state against decoherence.  Namely, if the state of interest is in subsystem $A$ and decoherence is stronger in subsystem $A$ than in $B$, then swapping protects against decoherence, as the state spends half its time in subsystem $B$ where there is less decoherence.  This is shown in Fig. \ref{fig:swap protects decay} where the swapping protocol plus collective decay $\gamma_-=0.1$, $\gamma_\text{int}=1$  (with the same system parameters as Fig. \ref{fig:Decoherence}) is compared to collective decay when $\gamma_\text{int}=\gamma_M=\gamma_J = 0$.      

\end{document}